# Reconnecting Citizens to Politics via Blockchain - Starting the Debate

## Uwe Serdült

*Centre for Democracy Studies Aarau (ZDA) at the University of Zurich, Switzerland, uwe.serdult@zda.uzh.ch; College of Information Science and Engineering, Ritsumeikan University, Japan, serdult@fc.ritsumei.ac.jp; Orcid [0000-0002-2383-3158].*

*Abstract: Elections are not the only but arguably one of the most important pillars for the proper functioning of liberal democracies. Recent evidence across the globe shows that it is not straightforward to conduct them in a free and fair manner. One constant concern is the role of money in politics, more specifically, election campaign financing. Frequent scandals are proof of the difficulties encountered with current approaches to tackle the issue. Suggestions on how to overcome the problem exist but seem difficult to implement. With the help of blockchain technology we might be able to make a step forward. A separate crypto currency specifically designed to pay for costs of political campaigning and advertising could be introduced. Admittedly, at this stage, there are many open questions. However, under the assumption that blockchain technology is here to stay, it is an idea that deserves further exploration.*

*Acknowledgement: A thank you note goes to participants of the Procivis Think Tank meeting, 21 September 2018, held at Trust Square, Zurich, Switzerland, for valuable inputs and discussions.*

## 1. Democratic Malaise

At the current state of affairs in the history of democracies across the globe, we are facing a paradoxical situation. On one hand we were able to observe the emergence and growth of the number of formal democracies over the last couple of decades, mainly in Latin America and Europe. On the other hand there seems to be a certain and deeply rooted dissatisfaction with politics in general, even in well-established polities of the so-called Western world. The disillusionment about politics can partly be related to a lack of performance legitimacy in the aftermaths of the most recent financial and economic crises but not exclusively. According to Crouch (2004), the deeper reason for this conundrum is the fact that many democracies have entered a post democratic phase. In post democracies formal requirements of a democratic polity as we define them – including basic political freedoms and rights, the rule of law and the separation of powers just to mention a few – are fully met. Elections take place in an orderly fashion and governmental forces do alter. However, politics has moved beyond so called first moments of democracy characterized by genuine mass participation, deliberation and a broad engagement of civil society. Even though elections are obviously being held, citizens now know they do not really matter much anymore. In post democracies boredom, frustration and disillusionment about politics are kicking in. With the knowledge of public relations experts as well as scientists, professionalized political forces such as politicians and lobbying groups have learned how to manipulate public opinion thus turning elections campaigns into a theatre orchestrated for the mass media. In order to find out what people want politicians started imitating show business and marketing techniques to increase their chances of getting elected or re-elected. Party platforms have become more superficial and less discriminatory, leaving the voters clueless about content and vulnerable to the personalization of politics. Important policies are being negotiated behind closed doors once elections are over, mainly between governments and big business. Corruption is not only a problem of the global South but represents a deeply entrenched phenomenon in politics in general. Citizens take part in decision-making only very unfrequently or not at all. Democracies thus tend to turn into empty procedural shells which can formally not be called undemocratic but leave many citizens – better educated than ever in human history – frustrated with politics, low respect for politicians and ultimately erode the legitimacy of democracies. Even if we were not fully convinced by that grim characterization and the statement that many countries entered a post-democratic stage (Merkel, 2014) we must concede that a couple of the described symptoms are difficult to reason away. Additionally, in most established democracies we observe a decline in electoral turnout accompanied by the rather bizarre situation that state authorities frequently have to admonish citizens to turn out and run actual campaigns for that.

Digital democracy, at least in the early stages, has been seen as a beacon of hope. However, the hope that the less powerful or financially less potent actors can have a say in politics with as little as an internet connection did not come true. It might have led to a further disillusionment instead: using the communication tools of the internet seems to require professional solutions, time and money, thus again favoring the larger, well established and well financed organizations. This is not to say that there is not the occasional exception which did not exist before such as a blog entry or a tweet causing a turn of developments or mobilizing the masses for a while. But this is clearly the exception. Also, the more radical approach of trying to circumvent dirty politics with the help of the internet all together, leading to a more direct and at the same time liquid democracy, does currently not seem to be a valuable alternative. Day-to-day politics caught up with such approaches quickly and we cannot see them taking root yet. Granted, more time might be needed but even then the looming question is whether liquid democracy would be capable of producing coherent policy output in the long run. A third option which is currently representing the most common and accepted track, is a digital democracy interpreted in the sense of bringing the citizens closer to the state and public administrations so that more efficient problem solutions can be found. This digital democracy approach currently seems to have the highest potential but also comes with some dangers. The closer citizens interact with the state the higher the danger for data misuse (for example in case of a change of government to one of a different color). In general, elements of digital democracy did not yet seem able to unleash its full constructive power, on the contrary, with micro-targeting of potential electors and increasing tendencies to influence voters on social media, also from abroad, we currently are faced with fighting the unwanted effects of the digital in politics.

## 2. Current Election Campaign Financing Rules and Suggested Reforms

Elections and election campaigns play a crucial role in democracies. They represent the high service transferring the powers from citizens to representatives and rulers for yet another term. It is also a time when through acts, procedures and symbolism democracies become visible and even tangible for the largest part of the electorate. Arguably, elections are not the only but certainly one of the most fundamental pillars for the proper functioning of liberal democracies. Unfortunately, evidence across the globe demonstrates how difficult it is to conduct them in a free and fair manner, even in more advanced liberal democracies. In particular, a constant concern is the role of money in politics, more specifically, money used for election campaign financing. Frequent scandals are proof of the difficulties encountered with current approaches to tackle the issue. As reported in a comprehensive, systematic and up to date analysis of political finance in comparative perspective (Norris et al., 2015) by the Electoral Integrity project and partners (Global Integrity, 2015) we are forced to conclude that the de facto situation often does not match with what regulation would prescribe. While a few countries score high on electoral integrity without detailed legal frameworks on how money can be used during election campaigns the opposite is also possible. In the majority of the covered countries the regulation in place does not affect third party actors such as political action committees, unions and certain not for profit organizations. The monitoring for direct and indirect financing of election campaigns furthermore shows that rules are often not very specific, that documentation is incomplete, incomprehensible and delivered late. Furthermore, in most countries public resources are being used for campaigns which makes it more difficult to quantify real costs and to track their use. The report comes to the quite devastating conclusion that the violation of rules is rather the norm. Only in four out of the 54 monitored countries did the authors not find any violations during the most recent elections.

Literature shows that there is a long political struggle to regulate election campaign donations. Typically, legislation would require campaigners, mostly political parties and candidates, to register and disclose donations they receive (Gilland Lutz & Hug, 2010). However, experience shows the futility of this approach. Having personally being able to check what kind of documentation is routinely handed in under the regime of some of the Swiss cantons with a disclosure rule for political donations (Serdült, 2010; Braun Binder 2015) it is not surprising to comprehend the conclusions of the upper mentioned international reports. Several contributions for a political party taken down in the name of the treasurer of the respective political party are not conducive to increasing trust in the system. Reporting duties for donations in the official gazette simply being ignored for years demonstrate that controls seem to be lax and the consequences eventually not severe in the case of infringements. The maximum fine could be well beyond the amounts received. The rules are sometimes also relatively easy to circumvent. Larger amounts of money can be split up to go beyond the allowed amount and thus stay under the radar screen. Contributions can be made to organizations not falling under the legislation. They can also originate from or be channeled to accounts abroad. In case campaign ad space is voluntarily given away for free because of ideological proximity between media owners and political party there is even no money trail at all. Occasionally, campaign financing scandals pop up, but they probably only represent the tip of the iceberg. In sum, current regulation approaches seem to lead to bureaucratic and largely ineffective solutions. However, if current campaign regulations have no positive effect on the fairness of elections better solutions should be sought. A first cursory review of potential remedies found in the literature reveals the following:

- Ayres and Bulow (1996) suggested a fund controlling for the identity of donors but then to relay the money to the receiver anonymously,
- a Council of Europe (2004) report put forward the idea to use vouchers instead of the national currency,
- Lessig (2011) proposed a reform of campaign financing allowing citizens to donate vouchers.

Whereas governance of any campaign financing regulation is going to stay key despite of the technology applied, a distributed approach involving not only public administrations or electoral management boards created to supervise all matters elections but the public in general might help to achieve a paradigm shift in the not so distant future.

## 3. A Blockchain Technology-based Approach

Thanks to distributed leadger technology - colloquially referred to as blockchains (Wattenhofer, 2019) - new options are now available, helping to combine and implement ideas such as the introduction of vouchers and partly anonymous donations in a more persistent way. Lessig's notion of paper vouchers can directly be reformulated in blockchain terminology terms as tokens. The logic behind the introduction of vouchers is that donations and campaign financing increase the risk of corruption and that one should try to extract those money flows from a market in which transactions are difficult to trace. With blockchain technology political campaigns can be tracked and financed by a crypto token owned by the people. Such crypto vouchers can have a duration and nominal value. They can be issued and distributed for every election or even for a complete legislature. In case there is a need, additional vouchers could be released. Each country or constituency could therefore create its own vouchers within a couple of hours. The suggested blockchain system would allow tracing all flows of the campaign crypto currency and to keep the total amount spent under control. However, the arguably more important innovation of the suggested approach is not only the technical aspect per se but the direct involvement of citizens. Every registered voter would henceforth have a small but direct stake in the electoral race. As a much welcomed side effect, interest in politics and therefore turnout might increase as well.

As a starting point for further reflection, token owners would, in the first place, have the following three options: they can pass tokens on to any other citizen, candidate or political group (donate), sell them in a market (trade) or even decide not to use them at all (abstain). For contested electoral races the price of a token could go up. National banks would be potential organizations capable to issue campaign tokens. They can hold them in their books just like any other currency. Last but not least, the most important point, all citizens are directly reconnected to politics by receiving tokens which define the total amount for campaign spending.

Interdisciplinary research to study technical, regulatory, economic as well as behavioral dynamics of such a blockchain-based campaign financing solution is of course much needed. The following research questions can serve as a preliminary guide for the development of future feasibility studies. The by no means exhaustive list of questions can be grouped into four domains:

1) Governance and legal: Which legal requirements apply to a cryptocurrency for political campaigns? In particular, which constraints are imposed by the principle of economic freedom and what requirements must be met with regard to ensuring electoral freedom? Which monetary law provisions would be necessary for the introduction for a separate cryptocurrency for political campaigns?
2) Technological: How should the tokens be designed? How can token donations stay anonymous but become public when reaching a certain threshold? Can secure and easy to use applications for individual use be envisaged?
3) Economic: How much would the introduction and administration of campaign tokens cost regarding administration as well as energy consumption? How can those costs be financed?
4) Behavioural: How do citizens react to the idea of creating tokens to finance political campaigns? Would they be willing to use them? Which patterns of token use can we observe?

## 4. Discussion

The suggested additional use case of blockchain technology in the public domain (Rizal Batubara et al., 2018) for a core aspect of any democracy worthy of its name has the potential to shed new light on one of its long looming conundrums. It provides for a prospective and optimistic way on how technology can be helpful for the design of a future (if not completely but increasingly) digital democracy. Through a transdisciplinary approach comprising legal, economic, technical and experimental elements, the proposal to create a decentral election token provides public authorities, politicians and society at large with an innovative template on how campaign financing could look like in the not so distant future. Furthermore, the prospective aspect of the proposal allows lawmakers to update themselves on the future challenges regarding the application of blockchain technology in democratic political systems.

Feasibility studies could help to cast light on opportunities and risks for the use of blockchain technology in the public domain. In that regard, Switzerland with its frequent referendum votes and elections on three state levels is a particularly well-suited field of experimentation. Referendum topics at stake do not necessarily and always follow the party lines. They can be cross-cutting. We can therefore expect campaign financing donations to reveal non-partisan patterns as well. However, whether citizens and other stakeholders would mainly donate along partisan lines, diverge from the expected pattern or in an act of self-interest prefer to cash in their allocated amount by selling the vouchers in a market (Fehr & Fischbacher, 2003) is an empirical question which should be addressed in studies. Regulation will most probably need to be local and study results will therefore not travel very well to other polities. However, all research along those lines will certainly have an international appeal and novelty. Feasibility studies need not be restricted to referendum votes such as in the Swiss case. On the contrary, the fact that Seattle in 2015 started an experiment making use of publicly funded 100 USD paper vouchers during an electoral race hints at the fact that the proposed, digitally enhanced campaign financing solution is not of a purely speculative nature and deserves the attention of researchers, politicians and civil society organizations alike.

**About the Author**

*Uwe Serdült*

Uwe Serdült occupies a dual position as a full professor at Ritsumeikan University, Japan, and a principle investigator in the Center for Democracy Studies Aarau (ZDA) at the University of Zurich, Switzerland. He is interested in research at the intersection of the social and information sciences, particularly in the field of e-government and digital democracy.